\documentclass[sigconf]{acmart}

\usepackage{algorithm,algpseudocode}

\usepackage{graphicx} 
\usepackage{graphics}
\usepackage{xspace}
\usepackage{booktabs,makecell,tabularx}
\usepackage{colortbl}
\usepackage{tcolorbox}
\tcbuselibrary{many}
\usepackage[table]{xcolor}

\usepackage{fontawesome5}

\usepackage{xcolor} 
\usepackage{enumitem}
\usepackage{pifont}
\usepackage{multirow}
\usepackage{cuted}

\usepackage{hyperref}
\hypersetup{
    colorlinks=true,
    linkcolor=black,
    anchorcolor=black,
    filecolor=black,      
    citecolor=black,
    urlcolor=black
}
\usepackage{array}
\usepackage[caption=false,font=normalsize,labelfont=sf,textfont=sf]{subfig}
\usepackage{textcomp}
\usepackage{stfloats}
\usepackage{url}
\usepackage{verbatim}
\usepackage{graphicx}

\newcommand{\rev}[1]{\textcolor{black}{#1}}

\usepackage{mdframed}

\usepackage{tcolorbox}

\definecolor{zirui}{RGB}{16,109,156}

\definecolor{myyellow}{HTML}{FFF2CC}
\definecolor{myblue}{RGB}{255,255,255}

\usepackage{tcolorbox}

 \newtcolorbox{mybox}[2][]
  {
    colback = white, colframe = black, fonttitle = \bfseries,
    colbacktitle = black, 
    title=#2,#1}

\DeclareRobustCommand{\mybox}[2][gray!20]{%
\begin{tcolorbox}[
        breakable,
        left=0pt,
        right=0pt,
        top=0pt,
        bottom=0pt,
        colback=#1,
        colframe=#1,
        width=\linewidth, 
        enlarge left by=0mm,
        boxsep=5pt,
        arc=0pt,outer arc=0pt,
        ]
        #2
\end{tcolorbox}
}

\usepackage{tcolorbox}

\usepackage{array}
\usepackage{booktabs}

\definecolor{migblue}{HTML}{1F4E79}
\definecolor{miglightblue}{HTML}{EEF5FB}
\definecolor{miggray}{HTML}{F6F7F9}
\definecolor{migborder}{HTML}{D6DCE5}
\definecolor{migredbg}{HTML}{FFF1F1}
\definecolor{miggreenbg}{HTML}{F0FAF2}
\definecolor{migred}{HTML}{B3261E}
\definecolor{miggreen}{HTML}{1B7F3A}

\newtcolorbox{migrationcard}[2]{
  enhanced,
  colback=white,
  colframe=migborder,
  boxrule=0.55pt,
  arc=2pt,
  left=5pt,
  right=5pt,
  top=5pt,
  bottom=5pt,
  title={\textcolor{migblue}{\textbf{#1}}\quad #2},
  coltitle=black,
  fonttitle=\bfseries\small,
  colbacktitle=miglightblue,
  attach boxed title to top left={xshift=0pt,yshift=-1pt},
  boxed title style={
    colback=miglightblue,
    colframe=miglightblue,
    sharp corners,
    boxrule=0pt
  }
}

\newtcolorbox{codebox}{
  colback=miggray,
  colframe=migborder,
  boxrule=0.35pt,
  arc=1.5pt,
  left=3pt,
  right=3pt,
  top=3pt,
  bottom=3pt
}

\copyrightyear{2026}
\acmYear{2026}
\setcopyright{cc}
\setcctype{by}
\acmConference[ASE '26]{Proceedings of the 41st IEEE/ACM International Conference on Automated Software Engineering}{October 12--16, 2026}{Munich, Germany}
\acmBooktitle{Proceedings of the 41st IEEE/ACM International Conference on Automated Software Engineering (ASE '26), October 12--16, 2026, Munich, Germany}
\acmDOI{10.1145/3832783.3834383}
\acmISBN{979-8-4007-2882-2/2026/10}

\begin{document}

\title{Assessing the Cross-Version Applicability of Java Library Vulnerability Exploits}

\author{Zirui Chen}
\orcid{0009-0004-6236-9150}
\affiliation{%
  \institution{The State Key Laboratory of Blockchain and Data Security, Zhejiang University}
   \city{Hangzhou}
  \country{China}
}
\email{chenzirui@zju.edu.cn}

\author{Qi Zhan}
\orcid{0000-0002-6800-1857}
\affiliation{%
  \institution{The State Key Laboratory of Blockchain and Data Security, Zhejiang University}
   \city{Hangzhou}
  \country{China}
}
\email{qizhan@zju.edu.cn}

\author{Jiayuan Zhou}
\orcid{0000-0002-5181-3146}
\affiliation{%
  \institution{Queen’s University}
   \city{Kingston}
  \country{Canada}
}
\email{jiayuan.zhou@queensu.ca}

\author{Xing Hu}
\orcid{0000-0003-0093-3292}
\affiliation{
  \institution{The State Key Laboratory of Blockchain and Data Security, Zhejiang University}
   \city{Hangzhou}
  \country{China}
}
\email{xinghu@zju.edu.cn}

\author{Xin Xia}
\authornote{Corresponding author}
\affiliation{
  \institution{The State Key Laboratory of Blockchain and Data Security, Zhejiang University}
  \institution{Hangzhou High-Tech Zone (Binjiang) Institute of Blockchain and Data Security}
   \city{Hangzhou}
  \country{China}
}

\email{xin.xia@acm.org}

\author{Xiaohu Yang}
\orcid{0000-0003-4111-4189}
\affiliation{%
  \institution{The State Key Laboratory of Blockchain and Data Security, Zhejiang University}
   \city{Hangzhou}
  \country{China}
}
\email{yangxh@zju.edu.cn}

\begin{abstract}

Open-source software supply chain security relies heavily on assessing affected versions of library vulnerabilities. While prior studies have leveraged exploits for verifying vulnerability-affected versions, they point out a key limitation that exploits are version-specific and cannot be directly applied across library versions. Despite being widely acknowledged, this limitation has not been systematically validated at scale, leaving the actual applicability of exploits across versions unexplored. To fill this gap, we conduct the first large-scale empirical study on exploit applicability across library versions. We construct a comprehensive dataset consisting of 259 exploits spanning 128 Java libraries and 28,150 historical versions, covering 61 CWEs that account for 76.3\% of vulnerabilities in Maven. 
Leveraging this dataset, we investigate the root causes of inconsistencies between exploit cross-version execution behaviors and ground truth, while also exploring strategies for exploit migration.

Our results (RQ1) show that, even without migration, exploits achieve 83.0\% recall and 99.3\% precision in identifying affected versions in Java, outperforming most widely used vulnerability databases and assessment tools.
Notably, this capability enables us to contribute more than 1,400 confirmed missing affected versions to the CPE dictionary.
We investigate the remaining exploit failures (RQ2) and find that they mainly stem from compatibility issues introduced by library evolution and changing environmental constraints. Based on these observations, we manually migrate exploits for 1,885 versions and distill a taxonomy of 10 strategies from these successful adaptation cases (RQ3), thereby increasing the overall recall to 96.1\%. Our findings underscore the cross-version potential of exploits and lay a foundation for future research.

\end{abstract}



\maketitle

\section{Introduction}

Open-source libraries are essential components of the modern software ecosystem~\cite{Zhang2024SymBisect,Zhang2025mitigation,cassel2025nodemedic, Huang2026Package, Huang2026Package1}, with more than 90\% of software relying on open-source components~\cite{Kula1, Markus1, Synopsys1, he2021migration}.
While such extensive reuse accelerates software development, it also introduces security risks~\cite{Bavota1, Kula1, Zhou2024Magneto, zirui2024exploiting,He2023Dependent,shuhan2025vul,Li2023downstream,Wu2026Oss}. A single vulnerability in a widely used Java library can propagate to millions of downstream projects~\cite{CVE-2021-44228}. To mitigate these risks, developers heavily rely on vulnerability databases (e.g., NVD) to check whether their library versions are affected~\cite{Nguyen2013Unreliability, Wu2024SCA}. However, recent studies have shown that these databases suffer from inaccuracies~\cite{Croft2023report, Dong2019Inconsistence, Jo2021Report}, posing significant challenges for security maintenance.

Vulnerability exploits play a crucial role in understanding the security issues~\cite{Wu2026PoC}. Prior studies have leveraged exploits to improve the reliability of affected-version identification by constructing ground truth for affected versions~\cite{Wu2024Vision} and validating whether the identified versions are truly vulnerable~\cite{Dong2019Inconsistence}. These studies point out a key limitation: exploits working for one version cannot always be directly reused in other versions~\cite{Dai2021Exploit, Wu2024Vision, Dong2019Inconsistence, Zhang2024SymBisect}. However, this limitation has not been validated at scale. For example, the evaluation by Dai et al.~\cite{Dai2021Exploit} covered only 470 affected versions.

To bridge this gap, we conduct the first large-scale empirical study to evaluate the exploit cross-version applicability. We establish the largest Java library vulnerability exploit dataset to date, comprising 259 exploits covering 128 libraries and 28,150 versions (14,378 versions are vulnerable). Our dataset spans 61 CWEs, which cover 76.3\% (4,495 out of 5,889) of vulnerabilities in Maven. Using this dataset, we investigate the following research questions (RQs):

\begin{itemize}[leftmargin=*]

\item {\textbf{RQ1 (Cross-Version Applicability): To what extent can disclosed exploits be reproduced across versions?}
In this RQ, we quantify cross-version applicability by executing disclosed exploits across all historical library versions to identify vulnerable releases. We compare the performance against \ding{182} ground truth, \ding{183} five widely-used vulnerability databases, and \ding{184} two SOTA tools. 
Our study reveals that disclosed exploits exhibit potential in cross-version applicability, achieving an 83.0\% recall and a 99.3\% precision.
}

\item {\textbf{RQ2 (Limiting Factors): What are the primary factors limiting the applicability of exploits?} 
We observe unexpected exploit behaviors in 2,447 affected versions and 5,538 unaffected versions. To understand these inconsistencies, we conduct a systematic analysis of the root causes through a card-sorting process. Our results show that breaking changes introduced during evolution, such as library refactoring, constitute the dominant barriers, leading to environment issues or symbol resolution failures. In addition, outdated validation logic often causes false alarms.}

\item {\textbf{RQ3 (Exploit Migration): To what extent and how can exploits be migrated across versions?} 
We construct a comprehensive benchmark of migrated exploits by manually adapting failed cases on affected versions identified in RQ2.
Based on this dataset, we derive a taxonomy of 10 exploit migration strategies that increase the recall of exploit-based assessment to 96.1\% and resolve most environmental and symbol resolution failures.
}

\end{itemize}


Based on our findings, we outline implications for stakeholders in the software supply chain. For vulnerability databases, we demonstrate the necessity of adopting exploits in affected version assessment, evidenced by our contribution of more than 1,400 confirmed CPE omissions. For downstream developers, we advocate for cross-referencing multiple data sources and incorporating environmental constraints (e.g., JDK versions) when assessing vulnerability impact. Finally, for researchers, we provide the largest benchmark for exploit migration, outline directions for implementing automated exploit migration, and highlight the remaining challenges. The main contributions of our paper are as follows: 
\begin{itemize}[leftmargin=*]

\item {We construct a dataset of 259 Java library vulnerability exploits, which is the largest dataset to the best of our knowledge. In addition, we annotate the affected versions for each vulnerability.}

\item {We conduct large-scale evaluations of exploit applicability across 28,150 versions. Our results show that exploits can identify 83.0\% of affected versions, outperforming most vulnerability databases.}

\item {We analyze factors that affect the applicability of exploits, propose 10 adaptation strategies, and migrate exploits for 1,885 (77.1\%) affected versions where disclosed exploits failed.}

\item {We propose implications to vulnerability databases, future researchers, and downstream developers for supporting maintenance and identifying affected versions. }

\end{itemize}


\section{Methodology}
\label{sec: method}

\begin{figure}[htbp] 
  \centering	
  \resizebox{\linewidth}{!}{
  \includegraphics[width=0.95\linewidth]{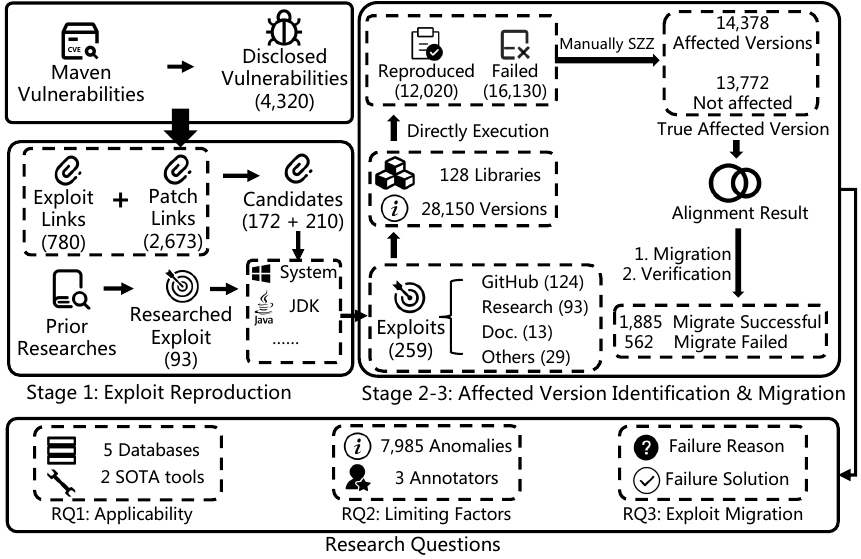}
  }
  \caption{Overall framework of our study.}
  \Description{Workflow of exploit collection, cross-version execution,
ground-truth construction, limiting-factor analysis, and exploit migration.}
  \label{fig: overview}
\end{figure}

An overview of our study is shown in Figure~\ref{fig: overview}. Previous studies have attempted to collect library vulnerabilities and their corresponding exploits~\cite{gao2025exploit,zirui2024exploiting}. However, these efforts lack a systematic collection criterion, resulting in small dataset sizes and low vulnerability coverage. To overcome this, we construct a large-scale library vulnerability exploit dataset to gain a comprehensive analysis of library vulnerability exploits. We implement a three-step process to answer our research questions:  \ding{182} Exploit Reproduction, \ding{183} Affected Version Identification, and \ding{184} Exploit Cross-version Migration.

\subsection{Exploit Collection}

We systematically collected disclosed exploits targeting the Java Maven ecosystem. Subsequently, we conducted a reproduction process to validate the executability of each collected exploit.

\subsubsection{Vulnerability Collection}

Our study targets vulnerabilities in the Maven ecosystem, which hosts over 50 million packages. 
All vulnerabilities in Maven are filtered according to a set of inclusion and exclusion criteria. We include a vulnerability if: \ding{182} The vulnerability belongs to the Maven Ecosystem. \ding{183} The vulnerability is disclosed before December 2024. We exclude a vulnerability if: \ding{182} The affected or patched version is not specified. \ding{183} The vulnerability has not been assigned a CVE number. Applying these criteria yields a dataset of 4,320 vulnerabilities affecting 1,981 libraries.

\subsubsection{Exploit Identification}

For each vulnerability, we collect candidate exploits from three sources to maximize coverage: 

\textbf{Reported Exploit.}
Vulnerability reports sometimes include reference links to known exploits. We extract links labeled as `exploit' from the NVD, resulting in 780 candidate links. Following prior practices~\cite{Mu2018Reproduce, Dai2021Exploit}, we filter these links using the following criteria:

\begin{itemize}[leftmargin=*]
    \item  \textit{Availability.} 
Links must be accessible during our collection process, leading to the exclusion of 74 exploit links.
\item  \textit{Relevance.} We retain exploits targeting Java library vulnerabilities. Exploits related to platforms (e.g., XWiki~\cite{XWiki}) or written in non-Java languages are omitted. Based on this, 216 exploits are removed because they targeted platforms rather than libraries, and 223 exploits are excluded due to unsupported languages.
\item \textit{Clarity.} A valid exploit must include both reproduction steps and details about the affected API and library. We discard links that lack either of these elements, excluding 37 exploits due to missing API information and 58 exploits due to unspecified inputs.

\end{itemize}

After filtering, we obtain 172 accessible links from various sources, including ExploitDB, Snyk, GitHub, and fuzzing frameworks such as OSS-Fuzz. These links correspond to 158 vulnerabilities.

\textbf{Researched Exploit.}
We further incorporate exploits collected in prior research~\cite{zirui2024exploiting, gao2025exploit, Wu2024Vision}, restricting our scope to those with valid CVE identifiers. These studies collectively include 132 Java library vulnerability exploits or vulnerability witness test cases. We exclude 39 cases because they verify the fix rather than triggering the vulnerability. As a result, we include 93 candidate exploits for 88 vulnerabilities from prior studies in our study.

\textbf{Constructed Exploit.} Inspired by Wu et al.~\cite{Wu2024Vision}, we analyze test cases embedded in vulnerability patches to construct exploits.  Based on the patch, we extract payloads to reproduce the vulnerable behavior, while ignoring tests that merely validate the fix (e.g., checking for an expected exception after applying the patch).  Based on these payloads, we reconstruct executable exploits to ensure that the vulnerability is genuinely triggered instead of validating the effectiveness of a patch.
We collect 2,673 patch references and ultimately select 210 candidate payloads covering 156 vulnerabilities following the same criteria of availability, relevance, and clarity.

\subsubsection{Exploit Reproduction}

We reproduce exploits in environments tailored to their specified requirements, such as library versions, Java versions, and operating systems. In cases where these requirements are unspecified, a default configuration is applied: Ubuntu 20.04, Java 8, and the latest unpatched library version. An exploit is considered reproduced if its runtime behavior aligns with the description provided in the corresponding vulnerability report~\cite{xue2024selfpico}.

We conduct a two-round reproduction process. First, a researcher with five years of experience in library vulnerabilities performs the initial reproduction. Subsequently, a second researcher independently verifies whether the reproduced exploits behave consistently with the reports. Despite our best efforts, some of the exploits could not be reproduced. In total, we reproduce \textbf{259 exploits} corresponding to \textbf{224 vulnerabilities} in 128 libraries after deduplication. In Section \ref{sec: summary}, we further analyze the collected vulnerabilities.

Surpassing the previously largest reproducible dataset~\cite{Wu2024Vision}, our dataset offers an advancement in both scale and quality. Specifically, \textbf{all exploits in our dataset are designed to trigger vulnerable behaviors}. Constructed over a rigorous three-month reproduction process, this work represents a comprehensive collection of reproducible Java library exploits to date.

\subsection{True Affected Version Identification}

Inspired by Chen et al.~\cite{chen2025far} and Bao et al.~\cite{Bao2022Vszz}, we devise a multi-stage process to identify affected versions across 28,150 versions.

\textbf{Stage 1: Fix Identification.} 
    For each vulnerability, we collect related information from multiple sources, including CVE descriptions, NVD reports, GitHub issues, and pull requests to identify the \textit{Fixing Commit} and associated vulnerable statements (i.e., the logic causing the security flaw).

\textbf{Stage 2: Inducing Commit Tracing.} 
    To determine the introduction version of the vulnerability, the annotators trace the commit history of the vulnerable statements to identify the \textit{Vulnerability-Inducing Commit}. 
    In complex scenarios involving code refactoring or file renaming, the annotators manually verify the logical equivalence of the code to ensure accurate tracing.

\textbf{Stage 3: Affected Version Identification.} 
    Simply mapping versions by release time is insufficient when facing multi-branch maintenance. 
    Therefore, we manually inspected the source code of the release artifacts to verify the presence of the vulnerability. 
    A version is labeled as vulnerable if it explicitly preserves the vulnerable logic and lacks the corresponding fix.

Two of the authors, each with five years of experience in software security, independently perform the identification. Their results show a high level of agreement, with a Cohen’s Kappa~\cite{cohen1960coefficient} of 0.911. Any disagreements are resolved with the third author. Using this method, we identify \textbf{14,378 true affected versions}. We spend two months constructing the ground truth.

\subsection{Exploit Execution and Migration}

\label{sec:migration}

\begin{figure}[htbp] 
  \centering	
  \resizebox{\linewidth}{!}{
  \includegraphics[width=1\linewidth]{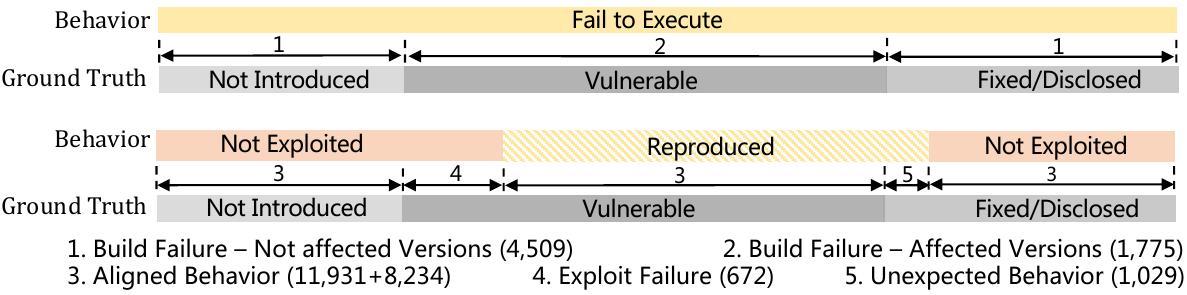}
  }
  \caption{Cross-version Execution Behavior of Exploits.}
  \Description{Comparison between exploit execution behavior and ground-truth
vulnerability ranges, including build failures, aligned behavior, exploit
failures, and unexpected behavior.}

  \label{fig: behavior}
\end{figure}

\textbf{Execution.} After establishing the ground truth, for each exploit we keep its originally reproduced environment and execute it across different library versions by only switching the vulnerable library version via \textit{mvn versions:use-dep-version}, without any additional changes. Exploits successfully trigger vulnerable behavior on 12,020 versions.
We align the exploit execution behavior in each version with our actual affected versions, mapping the observed behaviors to one of the given categories shown in Figure~\ref{fig: behavior}. \textbf{Build Failure} refers to cases where the exploit fails to execute on a given version. Among versions where the exploit executes successfully, \textbf{Aligned Behavior} indicates that the execution outcome aligns with the expectation. \textbf{Exploit Failure} denotes cases where the exploit fails to trigger the expected behavior on a version that is vulnerable in the true affected version set. \textbf{Unexpected Behavior} means capturing scenarios where the exploit remains effective or deviates from the normal behavior during execution.

By aligning the execution results with the ground-truth, we observe that in 11,931 vulnerable versions, the exploit behavior is consistent with the expected. However, in 2,447 vulnerable versions, the exploits fail to trigger directly, consisting of 1,775 build failures and 672 reproduction failures. 
In addition, we observe 1,029 cases where exploits exhibit unexpected behaviors on not vulnerable versions. 
We further discuss these discrepancies in Section ~\ref{sec: behavior}.

\textbf{Migration.} The 2,447 vulnerable versions for which exploits failed to reproduce constitute the primary subjects of our subsequent exploit migration analysis. While existing work has proposed automated techniques for Java API migration~\cite{Gao2021APIfix, Fruntke2025API}, we perform the exploit migration manually due to diverse reasons of reproduction failures and the requirement for payload adjustments. To ensure reliability, the migration process is conducted by two senior researchers, each possessing over five years of expertise in software security. Inspired by prior work on API migration~\cite{Fruntke2025API}, we conduct an exploit migration based on code changes. Specifically, for each failing version, we identify the nearest version where the exploit remains effective to serve as a reference version. We extract the source code diff files between the failing and reference versions, identify changes responsible for the failure, and derive the necessary adaptations accordingly. Two researchers attempt the migration independently, followed by a consolidation and cross-validation phase to ensure \textbf{the adapted exploits successfully reproduce the vulnerabilities} rather than accidentally trigger unrelated bugs~\cite{Zhang2024SymBisect}. We record the number of code lines modified and the time required for each migration. Ultimately, we successfully migrate exploits for 1,885 versions. However, 562 versions remain resistant to migration, due to reasons such as the complexity of modifying payloads generated by fuzzing. We provide a further analysis of these cases in Section ~\ref{sec:RQ3}.

\section{Data Summary}
\label{sec: summary}

Our dataset provides a comprehensive collection of Java library vulnerabilities with exploits, covering a wide range of libraries and CWEs, while including high-severity vulnerabilities.

\textbf{Library Categories.} We refer to the \textit{`Categories'} tags provided by \textit{MVNRepository}. Our dataset encompasses vulnerabilities affecting 128 libraries across 41 categories, ranging from testing frameworks and logging tools to JVM languages, HTTP clients, XML processors, and Object Serialization libraries. These categories cover diverse and essential components commonly integrated into modern software systems. Notably, 16 categories align with the most popular classifications, highlighting their widespread adoption.

\textbf{Library Usage.} Vulnerable libraries in our dataset account for over 400,000 usages in Maven. Among these, 57 libraries rank within the top 1,000 artifacts of Maven, emphasizing their critical role and the security risks. On average, each library is linked to 1.66 CVEs, with a median of 1 and a maximum of 35, the latter corresponding to the widely used \textit{jackson-databind} library.

\textbf{Vulnerability Severity.} Our dataset spans 61 CWEs, and 19 CWEs belong to the CWE Top 25, underscoring its relevance to prominent security challenges. We assess vulnerability severity based on the Common Vulnerability Scoring System (CVSS), which quantifies risk on a scale from 0 to 10. The average score in our collection is 7.75, with 62 vulnerabilities classified as `Critical' and 101 as `High', highlighting the significant severity.

\textbf{Vulnerability Coverage.} For the 5,889 Maven vulnerabilities disclosed up to 2025 and collected from GitHub Advisory, \textbf{76.3\% have at least one CWE covered by our dataset, while 68.9\% have all CWEs covered by our dataset}.  To evaluate the representativeness of our dataset across diverse vulnerability types, we adopt the \textit{CWE VIEW: Research Concepts}~\cite{CWEPillars}, which classifies weaknesses into abstract categories that reflect common structural or behavioral flaws across related CWEs (Pillars). By mapping each vulnerability’s CWE to its corresponding Pillar, we confirm that our dataset encompasses all major weakness categories. Details of the CWE categories are available in the replication package. 

\textbf{Exploit Source.} Our dataset comprises 259 vulnerability exploits. The majority of exploits originate from three key sources: 124 collected from GitHub (including 65 from issue discussions, 44 from commits and pull requests, 9 from security advisories, and 6 from miscellaneous links), 93 derived from prior research, and 13 extracted from documentation of vulnerable libraries.






















\section{RQ1: Cross-Version Applicability}
\label{sec: source}

We define \textit{Cross-Version Applicability} as the capability of a disclosed exploit to trigger the vulnerability on affected versions without modification. To quantify this capability, we compare the versions on which an exploit can be reproduced with versions reported by vulnerability databases and detected by SOTA tools. During data collection, we reproduce vulnerable behaviors on 12,020 versions via disclosed exploits (11,931 versions are vulnerable). This substantial dataset enables us to evaluate the cross-version applicability.

\subsection{Methodology} 

\subsubsection{Comparison with Databases.} We collect the affected library and version data from databases and compare these disclosures against the reproduced versions to assess the capability of exploits to successfully trigger vulnerabilities across different versions.

\textbf{Source Selection:} For each vulnerability, we extract the affected libraries and versions reported by five major data sources as of June 2025: NVD (CPE), GitHub Advisory (Affected versions), GitLab Advisory (Affected versions), Snyk (AFFECTS), and Veracode (Vulnerable range). Given the structural discrepancy of the affected library between CPE and Maven, we perform an additional mapping procedure to accurately align the disclosed CPE products with their corresponding artifacts in the Maven Central.
Importantly, we do not exclude unstable releases such as RC versions because they are available in Maven and have been adopted by users. 

\textbf{Evaluation Metrics:} Following prior studies~\cite{Wu2024Vision}, we evaluate performance using Precision and Recall. To avoid bias introduced by libraries with extensive version histories (e.g., \textit{jackson-databind}), we report \textit{Version-level Recall} and \textit{group-level macro Recall} over CVEs, libraries, and CWE types. Let $\mathcal{G}$ denote a set of groups (i.e., CVEs, libraries, or CWE types), and let $R_g$ denote the recall for a specific group $g \in \mathcal{G}$. Then:

\begin{equation*}
R_V = \frac{\sum TP}{\sum (TP + FN)}, \quad R_{\mathcal{G}} = \frac{1}{|\mathcal{G}|} \sum_{g \in \mathcal{G}} R_g, \quad \mathcal{G} \in \{CVE, Lib, CWE\}
\end{equation*}

To provide validation for the omissions identified by exploits, we selected NVD (CPE) as a representative data source and reported the omitted versions to the CPE Team. As a result, \textbf{more than 1,400 versions (spanning 19 CVEs) have been incorporated into the `Known Affected Software Configurations' field}.  The CPE Team confirmed via email that they ``\textit{have made the appropriate modifications to the configurations.}'' Regarding the reports that remain pending at the time of submission, we infer that the delay is likely due to the high workload, as the CPE Team has noted they are experiencing ``\textit{a large volume of CPE related inquiries}.'' We include all confirmed missing versions in our replication package.

\subsubsection{Comparison with SOTA Tools.} To the best of our knowledge, \texttt{Vision}~\cite{Wu2024Vision} represents the SOTA tool for identifying affected library versions in Java, reporting a precision of 0.91 and a recall of 0.94 in its dataset. Consequently, we select \texttt{Vision} and its underlying baseline, \texttt{V-SZZ}~\cite{Bao2022Vszz}, for our comparative evaluation. 
For each tool, we compare the affected versions it identifies against the versions on which the exploit reproduction succeeds. We further calculate precision and recall to evaluate the effectiveness of using exploits to identify affected versions without any modification.

To avoid introducing additional errors from identifying patches, we compare our execution results with the experimental results provided in the replication package of \texttt{Vision}~\cite{VisionReplication} overlapped by our datasets. The provided dataset includes results for 141 CVEs, with an overlap of 95 CVEs in our dataset, which we consider to be a representative subset for validation. This subset spans 76 libraries and comprises a total of 12,032 versions, consisting of 5,844 vulnerable versions and 6,188 unaffected versions.

\subsection{Assessment on Databases}

    


    

    
    
    

As illustrated in Table ~\ref{tab:alignment}, we evaluate the performance of data sources and exploits in assessing affected vulnerability versions. Based on this analysis, we draw the following conclusions:

\begin{table}[htbp]
  \centering
    \caption{Affected Version Identification Performance. }
  \label{tab:alignment}
  \scriptsize
  \setlength{\tabcolsep}{3pt}
  \resizebox{\columnwidth}{!}{%
  \begin{tabular}{c|c|ccccc}
    \toprule
    \multirow{2}{*}{\textbf{Metric}} & \multirow{2}{*}{\textbf{Exploit}} & \multicolumn{5}{c}{\textbf{Data Source}} \\
    \cline{3-7}
    & & \textbf{NVD} & \textbf{GitHub} & \textbf{GitLab} & \textbf{Snyk} & \textbf{Veracode}\rule{0pt}{2.6ex} \\
    \midrule
    
    \textbf{Identified} 
    & 11,931 & 10,843 & 11,164 & 10,036 & 13,156 & 11,581 \\

    \textbf{Recall (Version)} 
    & 83.0\%
    & 75.4\%
    & 77.7\%
    & 69.8\%
    & 91.5\%
    & 80.6\% \\

    \textbf{Recall (Group)} 
    & 81.4/79.9/78.9
    & 82.7/85.6/82.6
    & 76.1/73.5/84.2
    & 75.3/76.5/81.7
    & 89.7/88.3/88.3
    & 79.0/72.7/79.4 \\

    \midrule
    
    \textbf{Precision (Mis.)} 
    & 99.3\% (89) 
    & 90.9\% (1,082) 
    & 90.5\% (1,176) 
    & 88.9\% (1,255) 
    & 90.9\% (1,321) 
    & 93.9\% (752) \\

    \midrule
    
    \textbf{Exploit Unique} 
    & -- & 3,397 & 2,781 & 3,779 & 1,104 & 2,168 \\
    
    \textbf{Source Unique} 
    & -- & 2,309 & 2,014 & 1,884 & 2,329 & 1,820 \\
    
    \bottomrule
  \end{tabular}%
  }
\end{table}

\textbf{Exploits perform well in identifying affected versions against major data sources.} Overall, exploits achieve a version-level recall of 83.0\% in detecting affected versions. Notably, this performance surpasses four out of the five widely used databases, trailing only behind Snyk (91.5\%). Crucially, this performance is robust across library scales, as the CVE-level recall (81.4\%) and Library-level recall (79.9\%) closely align with the Version-level recall, demonstrating that the high applicability is observed across diverse vulnerabilities and libraries, rather than being dominated by a few libraries with extensive version histories. 

\textbf{Exploits demonstrate significantly lower false positive rates compared to common databases.} Among the 12,020 versions identified as vulnerable, exploits incurred only 89 false positives, achieving a precision of 99.3\%. In contrast, all compared data sources exhibit varying degrees of false positives. Even compared to Veracode, which achieves the highest precision, exploits demonstrate a 5.4\% improvement in precision. This increase in precision reduces the effort required to verify the impact.

\textbf{Exploits reveal five types of version omissions in data sources.} To evaluate the specific types of inaccuracies in vulnerability databases that can be uncovered by exploit execution, we focused our analysis on Snyk, which demonstrated the best performance among the evaluated databases. Even in the best-performing source, we find 1,104 affected versions confirmed through exploit reproduction are still overlooked, demonstrating that exploits can reveal affected versions missing from these sources. We categorized the identified omissions into five distinct types.


\begin{table*}[htbp]
\centering
\caption{Affected Version Identification Performance across CWE Types.}
\label{tab:cwe_results}
\scriptsize
\resizebox{\textwidth}{!}{
\begin{tabular}{lllcccccccccccc}
\toprule
\multirow{2}{*}{}
& \multirow{2}{*}{\textbf{Level}}
& \multirow{2}{*}{\textbf{Method}}
& \textbf{CWE-502}
& \textbf{CWE-611}
& \textbf{CWE-787}
& \textbf{CWE-22}
& \textbf{CWE-20}
& \textbf{CWE-770}
& \textbf{CWE-79}
& \textbf{CWE-94}
& \textbf{CWE-835}
& \textbf{CWE-121}
& \textbf{OTHERS}
& \multirow{2}{*}{\makecell{\textbf{OVERALL}\\\textbf{AVERAGE}}} \\
\cmidrule(lr){4-14}
&
&
& \#V.=48
& \#V.=20
& \#V.=19
& \#V.=18
& \#V.=12
& \#V.=11
& \#V.=11
& \#V.=9
& \#V.=8
& \#V.=6
& 51 CWEs
& \\
\midrule

\multirow{10}{*}{\textbf{Rec.}}
& \multirow{2}{*}{Version}
& Exploit
& \textbf{90.59\%}
& 80.37\%
& \textbf{93.69\%}
& \textbf{85.62\%}
& 76.13\%
& 73.93\%
& 53.44\%
& 57.95\%
& \textbf{89.52\%}
& 80.85\%
& 79.01\%
& 78.88\% \\
&
& Database
& 72.94\%
& \textbf{85.40\%}
& 81.94\%
& 71.68\%
& \textbf{91.36\%}
& \textbf{86.33\%}
& \textbf{73.70\%}
& \textbf{87.39\%}
& 85.52\%
& \textbf{98.09\%}
& \textbf{83.21\%}
& \textbf{83.25\%} \\
\cmidrule(lr){2-15}

& \multirow{2}{*}{CVE}
& Exploit
& \textbf{83.54\%}
& \textbf{81.62\%}
& \textbf{91.61\%}
& \textbf{85.46\%}
& \textbf{84.62\%}
& 67.43\%
& 49.73\%
& 70.18\%
& 86.12\%
& 80.92\%
& 79.62\%
& 79.37\% \\
&
& Database
& 77.49\%
& 78.37\%
& 82.93\%
& 77.08\%
& 80.83\%
& \textbf{79.57\%}
& \textbf{76.50\%}
& \textbf{81.96\%}
& \textbf{91.85\%}
& \textbf{98.32\%}
& \textbf{83.94\%}
& \textbf{83.70\%} \\
\cmidrule(lr){2-15}

& \multirow{2}{*}{Library}
& Exploit
& 75.54\%
& \textbf{82.49\%}
& \textbf{93.50\%}
& \textbf{86.23\%}
& \textbf{84.62\%}
& 65.82\%
& 54.85\%
& \textbf{79.13\%}
& 78.04\%
& 81.03\%
& 80.13\%
& 79.80\% \\
&
& Database
& \textbf{78.74\%}
& 77.92\%
& 78.59\%
& 75.83\%
& 80.83\%
& \textbf{75.07\%}
& \textbf{74.12\%}
& 77.13\%
& \textbf{86.96\%}
& \textbf{97.84\%}
& \textbf{83.54\%}
& \textbf{83.01\%} \\
\cmidrule(lr){2-15}

& \multirow{2}{*}{CWE}
& Exploit
& 73.73\%
& 67.18\%
& \textbf{91.41\%}
& \textbf{96.41\%}
& \textbf{95.69\%}
& 39.91\%
& 55.42\%
& 45.02\%
& \textbf{94.76\%}
& 81.51\%
& 78.49\%
& 77.77\% \\
&
& Database
& \textbf{93.40\%}
& \textbf{83.23\%}
& 80.02\%
& 78.63\%
& 83.07\%
& \textbf{93.16\%}
& \textbf{80.81\%}
& \textbf{95.73\%}
& 80.94\%
& \textbf{98.01\%}
& \textbf{84.15\%}
& \textbf{84.57\%} \\
\midrule

\multirow{10}{*}{\textbf{Prec.}}
& \multirow{2}{*}{Version}
& Exploit
& \textbf{99.54\%}
& \textbf{100.00\%}
& \textbf{98.22\%}
& \textbf{100.00\%}
& \textbf{100.00\%}
& \textbf{100.00\%}
& 89.56\%
& \textbf{100.00\%}
& \textbf{100.00\%}
& \textbf{100.00\%}
& \textbf{99.79\%}
& \textbf{99.61\%} \\
&
& Database
& 97.41\%
& 98.75\%
& 97.89\%
& 85.95\%
& 96.83\%
& 94.51\%
& \textbf{97.55\%}
& 99.77\%
& 62.77\%
& \textbf{100.00\%}
& 86.34\%
& 87.46\% \\
\cmidrule(lr){2-15}

& \multirow{2}{*}{CVE}
& Exploit
& \textbf{98.44\%}
& \textbf{100.00\%}
& \textbf{99.61\%}
& \textbf{100.00\%}
& \textbf{100.00\%}
& \textbf{100.00\%}
& 86.60\%
& \textbf{100.00\%}
& \textbf{100.00\%}
& \textbf{100.00\%}
& \textbf{99.85\%}
& \textbf{99.62\%} \\
&
& Database
& 96.36\%
& 81.60\%
& 87.12\%
& 84.06\%
& 87.03\%
& 82.04\%
& \textbf{93.00\%}
& 84.23\%
& 89.00\%
& \textbf{100.00\%}
& 84.86\%
& 85.45\% \\
\cmidrule(lr){2-15}

& \multirow{2}{*}{Library}
& Exploit
& \textbf{91.67\%}
& \textbf{100.00\%}
& \textbf{99.71\%}
& \textbf{100.00\%}
& \textbf{100.00\%}
& \textbf{100.00\%}
& \textbf{94.54\%}
& \textbf{100.00\%}
& \textbf{100.00\%}
& \textbf{100.00\%}
& \textbf{99.82\%}
& \textbf{99.62\%} \\
&
& Database
& 87.82\%
& 81.50\%
& 83.54\%
& 80.55\%
& 87.03\%
& 78.16\%
& 92.25\%
& 79.73\%
& 82.40\%
& \textbf{100.00\%}
& 84.52\%
& 84.65\% \\
\cmidrule(lr){2-15}

& \multirow{2}{*}{CWE}
& Exploit
& \textbf{99.97\%}
& \textbf{100.00\%}
& \textbf{99.41\%}
& \textbf{100.00\%}
& \textbf{100.00\%}
& \textbf{100.00\%}
& \textbf{96.52\%}
& \textbf{100.00\%}
& \textbf{100.00\%}
& \textbf{100.00\%}
& \textbf{99.81\%}
& \textbf{99.77\%} \\
&
& Database
& 96.82\%
& 98.00\%
& 83.36\%
& 83.96\%
& 87.08\%
& 94.34\%
& 92.31\%
& 99.94\%
& 71.39\%
& \textbf{100.00\%}
& 87.16\%
& 87.74\% \\
\bottomrule
\end{tabular}
}
\end{table*}

\begin{enumerate}[leftmargin=*]

   \item \textbf{Library Misreported} (350 Versions, 10 Vulnerabilities): This category represents cases where the database incorrectly identifies the vulnerable library and reports another library, such as the parent or sibling module of the vulnerable library. For example, CVE-2023-39010 affects \textit{org.boofcv:boofcv-core}, yet it is reported as \textit{org.boofcv:ip}. Similar mismatches are observed in JLine (reporting \textit{jline-parent} instead of \textit{jline-groovy}) and Jackson (reporting \textit{toml} instead of \textit{text} format).

    \item \textbf{Branch Overlooked} (411 Versions, 30 Vulnerabilities): This category represents omissions of branches affected by the vulnerability that are not covered by the data sources. 
    For example, in CVE-2020-24616, Snyk only included versions before \textit{2.9.10.6} of \textit{jackson-databind}, whereas the exploit is successfully reproduced on several versions in the \textit{2.10} and \textit{2.11} branches. However, our source-code analysis shows that the vulnerable logic persists in the 2.10 and 2.11 branches until it is fixed in versions \textit{2.10.5.1} and \textit{2.11.3}, respectively. Therefore, versions preceding these fixes in both branches are affected but omitted by Snyk.
    
    \item \textbf{Incorrect Introduce Range} (144 Versions, 5 Vulnerabilities): This category refers to affected versions earlier than the disclosed version that are omitted due to incorrect annotation of the introduced version range. 
    For example, in CVE-2019-10086, Snyk identifies the range as \textit{commons-beanutils:[1.9.2,1.9.4)}, yet exploits successfully reproduce the vulnerability in versions as early as \textit{[1.0,1.9.1]}. Similarly, for CVE-2022-23082, \textit{curekit:1.0.0} is omitted despite being vulnerable.

    \item \textbf{Incorrect Fix Range} (19 Versions, 7 Vulnerabilities): Similarly to \textit{Incorrect Introduce Range}, this category refers to affected versions higher than the disclosed version that are omitted because of incorrect annotation of the fixed version range. 
    For example, in CVE-2021-43795, Snyk excludes \textit{armeria:[1.12.0,1.13.3]} from the affected versions, whereas the exploit can be reproduced.

     \item \textbf{Unstable Overlooked} (180 Versions, 43 Vulnerabilities): This category represents omissions of unstable or pre-release versions (e.g., alpha, beta, RC). For example, in CVE-2018-19360, versions \textit{2.7.0-rc1} to \textit{2.7.0-rc3} successfully reproduce the vulnerability with the disclosed exploit, but are omitted from Snyk. Despite their pre-release status, these versions serve usage counts exceeding 100 in total on Maven Central. Thus, it is essential to include these versions during vulnerability assessments.
     
\end{enumerate}

\textbf{Exploits fall short in detecting affected versions, averaging 2,071.2 missed versions across data sources.} Although exploit cross-version execution outperforms four databases in detected affected version count, it does not fully replace them. The  `Source Unique' row in Table  ~\ref{tab:alignment} indicates the number of affected versions identified by the databases but missed by exploits, averaging 2,071.2 missed versions per source. \textbf{Exploit-based identification does not outperform existing databases in terms of overall CWE-level recall.} As illustrated in Table ~\ref{tab:cwe_results}, although it outperforms databases on several types, including CWE-502, CWE-22, and CWE-787, it underperforms on several categories such as CWE-79.

\mybox{
\textbf{Finding 1}: \rev{
 Our study reveals that disclosed exploits exhibit \textbf{remarkable cross-version applicability} compared to databases and tools. They identify \textbf{11,931} affected versions (83.0\% recall) with high precision and leave room for further improvement.
  }
}

\subsection{Assessment on SOTA tools}

In this section, we evaluate whether the effectiveness of identifying affected versions via cross-version exploit execution is better than static analysis approaches (e.g., \texttt{Vision} and \texttt{V-SZZ}). As illustrated in Table ~\ref{tab:comparison}, by comparing exploits with those identified by SOTA methods, we draw the following observations:

\begin{table*}[htbp]
\centering
\caption{Affected Version Identification Results of SOTA Approaches.}
\label{tab:comparison}
\small
\resizebox{0.8\textwidth}{!}{
\begin{tabular}{lcccccccccccc}
\toprule
\multirow{2}{*}{\textbf{Method}}
& \multicolumn{4}{c}{\textbf{Results}}
& \multicolumn{2}{c}{\textbf{Version-level}}
& \multicolumn{2}{c}{\textbf{CVE-level}}
& \multicolumn{2}{c}{\textbf{Library-level}}
& \multicolumn{2}{c}{\textbf{CWE-level}} \\
\cmidrule(lr){2-5}
\cmidrule(lr){6-7}
\cmidrule(lr){8-9}
\cmidrule(lr){10-11}
\cmidrule(lr){12-13}
& \textbf{TP}
& \textbf{FP}
& \textbf{FN}
& \textbf{Unique}
& \textbf{Prec.}
& \textbf{Rec.}
& \textbf{Prec.}
& \textbf{Rec.}
& \textbf{Prec.}
& \textbf{Rec.}
& \textbf{Prec.}
& \textbf{Rec.} \\
\midrule
Exploit
& 4,584
& 10
& 1,260
& 422
& \textbf{99.78\%}
& 78.44\%
& \textbf{96.94\%}
& \textbf{81.93\%}
& \textbf{98.80\%}
& 82.44\%
& \textbf{99.82\%}
& \textbf{82.94\%} \\

\texttt{Vision}
& 4,624
& 755
& 1,220
& 318
& 85.96\%
& \textbf{79.12\%}
& 83.92\%
& 79.21\%
& 85.90\%
& \textbf{82.57\%}
& 88.05\%
& 81.97\% \\

\texttt{V-SZZ}
& 2,893
& 869
& 2,951
& 169
& 76.90\%
& 49.50\%
& 51.34\%
& 45.39\%
& 53.70\%
& 46.71\%
& 62.43\%
& 47.02\% \\
\bottomrule
\end{tabular}
}
\end{table*}

\textbf{Exploit execution achieves substantially higher precision while maintaining recall comparable to the best-performing SOTA approach.}
Exploit execution identifies affected versions with a precision of 99.78\%, producing only 10 false positives. In contrast, \texttt{Vision} and \texttt{V-SZZ} produce 755 and 869 false positives, corresponding to precisions of 85.96\% and 76.90\%, respectively. Moreover, exploit execution achieves a recall of 78.44\%, only 0.7 percentage points lower than \texttt{Vision} (79.12\%) and substantially higher than \texttt{V-SZZ} (49.50\%). These results indicate that cross-version exploit execution can substantially reduce false positives while preserving competitive coverage of affected versions.

\textbf{Exploit execution complements SOTA approaches by uncovering additional affected versions.}
Cross-version exploit execution identifies 422 affected versions that are missed by both \texttt{Vision} and \texttt{V-SZZ}. These uniquely identified versions show that exploit execution provides complementary evidence beyond patch- and history-based analysis. Therefore, combining exploit execution with existing static approaches may enable more comprehensive and reliable affected-version identification.

Compared with these tools, we find that exploits perform better mainly in two scenarios. First, for vulnerabilities spanning multiple branches, patch-based methods tend to cover only branches near the patch. For example, exploits recover 118 affected versions missed by \texttt{Vision} for CVE-2022-42004 in \textit{jackson-databind}.
Second, exploit-based evaluation is effective when patch-based methods incorrectly identify the vulnerability introduced range. For example, for CVE-2023-43642 in \textit{snappy-java}, \texttt{Vision} identifies 1.0.4.1 as the introduction version and therefore misses earlier affected versions, while the exploit reproduces the vulnerability on these versions.

\mybox{
\textbf{Finding 2}: \rev{
    While maintaining a recall comparable to SOTA tools, exploits achieve significantly higher precision and identify affected versions that SOTA tools fail to detect.
    }
}

\section{RQ2: Limiting Factors}
\label{sec: behavior}

While our results confirm the cross-version applicability, exploits still face compatibility issues in 7,985 versions. In this RQ, we systematically analyze the factors contributing to this issue. Specifically, we conduct a card sorting analysis on versions whose exploit behaviors do not align with the expectation of true affected versions, aiming to uncover the reasons behind these discrepancies.

\subsection{Methodology}

\subsubsection{Labeling}
We systematically analyze execution behaviors using Hybrid Card Sorting~\cite{cardsorting}, a widely adopted classification technique. The procedure, including cross-validation, is performed by the first three authors, each possessing over three years of experience in open-source software security. We divide the vulnerabilities into two equal parts, with each of the first two authors independently responsible for one part. For behaviors that are not classified as \textbf{aligned behavior} in Figure ~\ref{fig: behavior}, the annotators abstract the causes of the unexpected behavior into concise, card-like descriptors. This process initially produces 34 distinct cards, which are later merged into a final set of 13 cards through collaborative discussion.

\subsubsection{Cross-Review}
An independent annotator, who does not participate in the initial labeling, is invited to validate the results. From each annotator’s part, we randomly select 20 vulnerability samples and assign them to another two annotators who are not involved in the labeling task of the vulnerabilities. If both annotators agree with the initial label, the label is confirmed. In cases of disagreement, the two reviewers resolve the conflict through discussion to reach a consensus. The entire annotation and cross-review process takes the first three authors two months to complete.

\subsection{Overview of Applicability Assessment}

\begin{figure*}[htbp] 
  \centering	
  \includegraphics[width=0.85\linewidth]{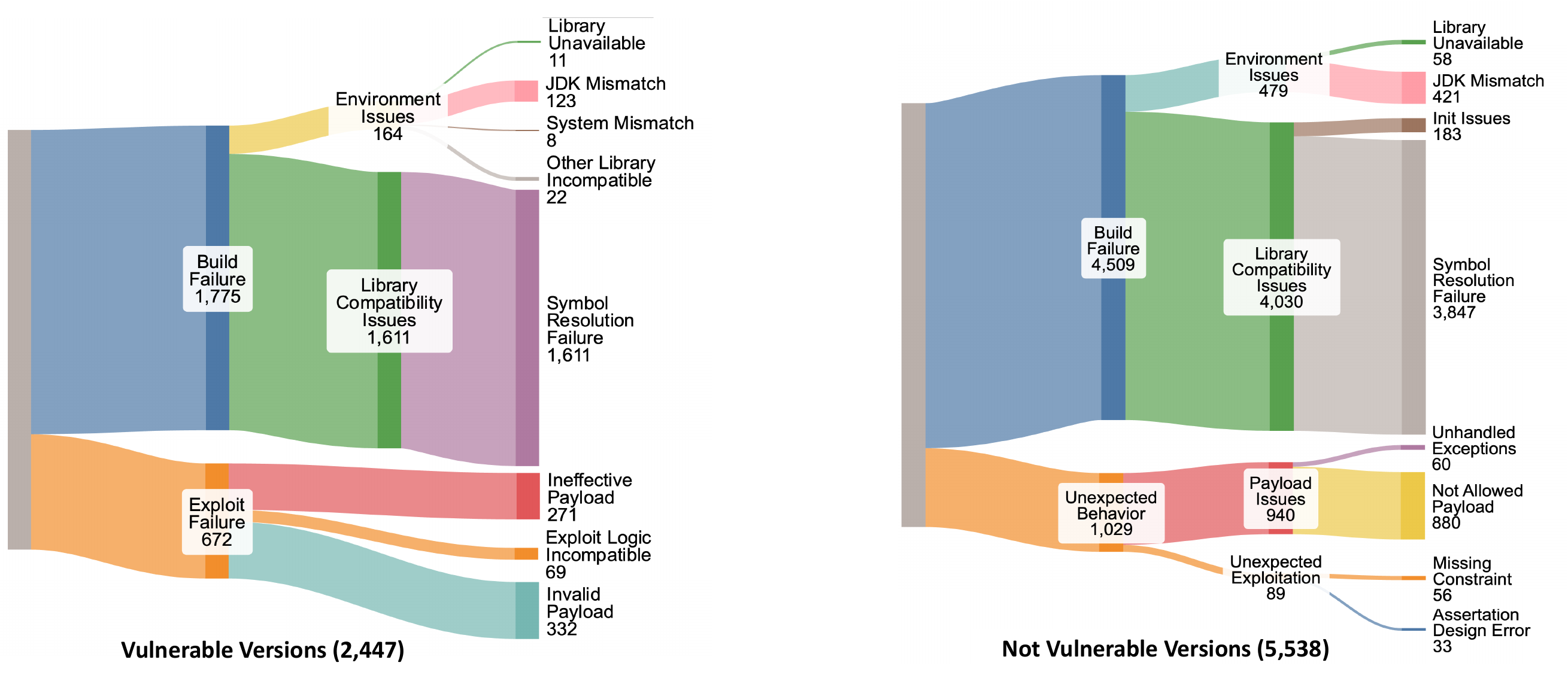}
  \caption{ Reasons for Unexpected Behaviors.}
  \Description{Sankey diagram showing the causes of build failures, exploit
failures, and unexpected behaviors on vulnerable and non-vulnerable versions.}
  \label{fig: summary}
\end{figure*}

Building upon the work of Zhang et al. ~\cite{Zhang2025mitigation}, we introduce a hierarchical taxonomy to analyze the underlying causes of diverse execution behaviors, as depicted in Figure~\ref{fig: summary}.  This taxonomy consists of different levels: the first level delineates high-level categories of unexpected behaviors; the second level refines these behaviors into subcategories; and the other levels enumerate specific causes associated with each subcategory. Levels may be left unpopulated when no suitable subcategories or concrete causes can be identified. This hierarchical framework facilitates a systematic investigation into why nearly 25\% of exploit executions deviate from expected.

The challenges of cross-version applicability stem from three factors: Build Failure, Exploit Failure, and Unexpected Behavior. \textbf{Build Failure} manifests in both vulnerable (1,775 versions) and non-vulnerable  (4,509 versions) versions. These failures are classified into two categories: incompatibility issues arising from different library versions and environment issues due to specific configurations. The causes of \textbf{Exploit Failure} (672 versions) are grouped into three categories: flaws or incompleteness in the payload design that prevent reproduction, payload non-compliance with version-specific input validations, and incompatibilities between the exploit verification logic and the runtime behavior of the library. Finally, \textbf{Unexpected Behavior} (1,029 versions) can be divided into two groups: cases where the payload is intercepted or encounters exceptions on safe versions leading to unexpected results, and cases where the exploit still succeeds on safe versions due to inadequate constraint or defective vulnerability verification logic. In the following sections, we provide a detailed breakdown of each category and present concrete examples to illustrate how these factors lead to reproduction failures.

\subsection{Build Failure} 

Build Failure refers to cases where the exploit fails to execute after changing the library version, which occur in 1,775 cases and 4,509 cases in both vulnerable and non-vulnerable versions. 

The vast majority (5,641 cases in total, 89.8\%) are attributed to \textbf{Library Compatibility Issues}, where breaking changes introduced during library evolution disrupt the execution of the exploit.

\begin{itemize}[leftmargin=*]

\item{\textsc{Symbol Resolution Failure}: Most compatibility issues manifest as `not found', such as \textsc{Class not Found} (3,578), \textsc{Method not found} (1,390), \textsc{Constructor not Found} (128), \textsc{Package not Found} (190) errors. For example, executing the exploit for CVE-2018-18893 on \textit{jinjava:2.4.1} triggers a \textit{`cannot find symbol'} error because the method is named \textit{getErrors()} in the targeted version.}

\item{ \textsc{Init Issue}: This issue arises from internal logic changes in updated versions and leads to not initialized or could not be instantiated errors. For example, in \textit{spring-webmvc:5.1.0.RELEASE}, the \textit{ResourceResolverChain} must be initialized before executing the exploit, which is not required when designing the exploit. }

\end{itemize}

The remaining 643 failures (10.2\%) are attributed to \textbf{Environment Issues}, which can be further categorized into four types:

\begin{itemize}[leftmargin=*]

\item{\textsc{JDK Mismatch} (544): Failures caused by updated libraries requiring a higher JDK version than the current environment. For instance, our collected exploit for CVE-2024-22257 relies on JDK 11, whereas \textit{spring-security-core:[6.0.0,)} requires JDK 17. This incompatibility leads to a version mismatch error indicating that \textit{`class file has wrong version 61.0, should be 55.0'}}.

\item{\textsc{Library Unavailable} (69): After changing the library version in \textit{pom.xml}, some specific JARs cannot be retrieved from the Maven Central Repository, leading to build failures. For instance, in the case of CVE-2013-2055, the library \textit{wicket-core:6.1.0} depends on \textit{wicket-util:6.1.0}. However, since this dependency is missing, the build process fails with a \textit{`Could not transfer artifact'} error.}

\item{\textsc{Other Library Incompatibility} (22): Some exploits depend on specific third-party environments (e.g., logging frameworks). These environments are unusable in the modified library versions. For example, reusing the logging configuration of \textit{esapi:2.2.1.0} in lower versions leads to a \textit{`LogFactory not in class path'} issue.}

\item{\textsc{System Mismatch} (8): Although the exploit is effective on the disclosed version, the updated library modifies the underlying implementations related to the operating system. 
For instance, \textit{`Error looking up function Java\_java\_lang\_UNIXProcess\_forkAndExec'} 
indicates that the updated library only supports Linux.}

\end{itemize}

\subsection{Exploit Failure}


As illustrated in Figure~\ref{fig: summary}, exploit Failure refers to cases where an exploit executes successfully but fails to trigger the vulnerable behavior, which accounts for 672 versions (27.5\%) of false negatives.

\textbf{Invalid Payload} refers to exploits that failed to reproduce in 332 versions because \ding{182} payload syntax accepted by the originally disclosed vulnerable version is rejected by the target version, or \ding{183} the required gadgets are absent from the target version. For example, the exploit for CVE-2023-34620 fails against \textit{hjson:1.1.2} because this version performs stricter syntax validation and rejects the unquoted key names used in the original payload.
Similarly, the exploit for CVE-2023-34455 fails on \textit{snappy-java:1.1.2}. The payload is compressed with an unsupported \textit{codec}.

\textbf{Ineffective Payload} denotes cases (271 Versions)  where the payload is compliant with syntax rules but failed to trigger the vulnerability caused by the reliance of payloads on specific internal implementation details that are altered across versions. For instance, changes in deserialization logic can render the payload for CVE-2020-9546 invalid as it uses types that are no longer supported, manifesting as messages like \textit{`Can not find a deserializer for non-concrete Collection type'}. Similarly, the payload for CVE-2018-21234 implicitly depends on the execution sequence of serialized fields. If the underlying \textit{jodd-json} implementation is refactored and fails to preserve field order, the specific execution is disrupted.

\textbf{Incompatible Exploit Logic} refers to cases (69 Versions) where the validation logic in the disclosed exploit is too specific and fails to account for variations in execution behavior across different versions. For instance, the exploit for CVE-2016-7051 validates vulnerability presence by checking for the specific error message \textit{`Connection Refused...'} when accessing a non-existent server. However, in \textit{jackson-dataformat-xml} versions \textit{[2.0.0,2.7.7]}, this exception is not thrown under the same conditions.

\subsection{Unexpected Behavior} Unexpected Behavior refers to exploits triggering unexpected exceptions or reproducing the vulnerability on non-vulnerable versions. As shown in Figure~\ref{fig: summary}, we identify 1,029 such cases.

The majority of cases (940 Versions, 91.4\%) are attributed to \textbf{payload issues}, where the payload employed by the exploit is invalid in non-affected versions, leading to anomalous behavior.

\begin{itemize}[leftmargin=*]

\item{\textsc{Not Allowed Payload} (880 Versions) refers to cases where the payload is intercepted by validation logic introduced in the security patch. Since the original exploit design does not account for this mitigation, the test outcome manifests as a failure. However, the output triggers an exception indicating that the input is blocked (e.g., \textit{prevented for security reasons} in \textit{jackson-databind}).}

\item{\textsc{Unhandled Exceptions} (60 Versions) refers to cases in which exploits exhibit newly introduced, potentially harmful behaviors in patched versions. For example, in CVE-2018-1324, although version 1.16 is marked as patched, executing the exploit on this version triggers an unreported \textit{ArrayIndexOutOfBoundsException}.}

\end{itemize}

The remaining 89 versions are attributed to \textbf{unexpected exploitation}, which means the exploit assertion is being satisfied even on patched versions. This leads to false positives and compromises the precision of exploit-based affected version identification.

\begin{itemize}[leftmargin=*]

\item{\textsc{Missing Constraints} (56 Versions) refers to cases where the exploit fails to incorporate the security constraints introduced by the patch. In these scenarios, the mitigation relies on developers to enable new restrictions, which the original exploit ignores. A typical example is CVE-2022-45688, where the fix introduces a secure usage pattern via \textit{XML.toJSONObject(s, ParseConfig.of().setMaxNestingDepth(512))}. However, the exploit continues to invoke the unconstrained interface \textit{XML.toJSONObject(s)}. Consequently, the exploit remains effective because the secure configuration is not adopted, rather than the version being vulnerable, causing false positives. Similar cases include CVE-2018-21234, CVE-2021-43859, and CVE-2022-22885.}

\item{\textsc{Assertion Design Error} (33 Versions) refers to exploit failures resulting from outdated exploit validation logic. For instance, the exploit for CVE-2019-16869 validates the presence of an \textit{IllegalArgumentException}; however, this exception was refactored to \textit{InvalidLineSeparatorException} in \textit{netty-codec-http:[4.1.125.Final:)}, leading to an invalid assertion failure. Similarly, modifications to the input serializer in \textit{AntiSamy} caused the exploit ineffective.}

\end{itemize}

\mybox{
\textbf{Finding 3}: \rev{
    The primary influences on cross-version applicability are library compatibility issues and environmental mismatches. Additionally, payload failures across different versions frequently result in false negatives. Finally, outdated validation logic always leads to false positives.
    }
}

\section{RQ3: Exploit Migration}
\label{sec:RQ3}

To understand the challenges in exploit migration and summarize effective strategies, we conducted a qualitative analysis of our manual migration process. As illustrated in Section \ref{sec:migration}, out of the 2,447 false negative cases, we successfully migrated \textbf{1,885} versions. In total, the migration process required modifying 30,943 lines of code, providing a rich basis for analyzing common migration strategies.

\subsection{Methodology}

For each failure, we first migrate the exploit to one target version and then evaluate the migrated exploit on neighboring failed versions, since the same migrated exploit may work across nearby versions. If it succeeds, no additional migration is needed for those versions, and we attribute the same modification effort for the corresponding target version. In total, handling the 1,885 versions takes 7,805 minutes, including 2,522 person-minutes for migration and 5,283 person-minutes for evaluating neighboring versions. Then we adopted a three-step \textit{Open Coding} approach~\cite{corbin2014basics} to categorize the adaptation actions during our manual migration:

\begin{itemize}[leftmargin=*]
    \item \textbf{Diff Generation.} 
    For each successfully migrated case, we collect \ding{182} the environment configurations (e.g., JDK versions, dependency trees) and \ding{183} the exploit code. We generate textual diffs between the original failing exploits and the migrated working exploits to highlight the necessary modifications.
    
    \item \textbf{Labeling and Concept Extraction.}
    The first two authors independently inspect the mapping between the \textit{limiting factors} (e.g., Symbol Resolution Failure) and the \textit{adaptation actions} (e.g., ``Updated API call''). 
    Based on this, they assign labels to each migration case (e.g., ``JDK Version Adaptation'', ``API Renaming''). 
    
    \item \textbf{Step 3: Taxonomy Construction.}
    We employ an iterative process to group the initial labels into high-level strategies. 
    The authors discuss and merge similar adaptation patterns until a consensus is reached, resulting in 10 migration strategies.
\end{itemize}

\subsection{Results}

Based on the root causes of the inconsistency on vulnerable versions (\textit{Environment Issues}, \textit{Library Incompatibility Issues}, and \textit{Exploit Failure}), we conduct a further analysis of the corresponding migration strategies distilled from our manual migration.
\subsubsection{Environment Issues}

Among the 164 environmental issues, we successfully migrate \textbf{154} of them ($93.9\%$), indicating that this category of failures is highly resolvable.

\begin{itemize}[leftmargin=*]

\item \textbf{S1: Finding Nearest Library Version.} 
We successfully resolve 7 of the 11 \textsc{Library Unavailable} cases using this strategy. If the missing library is not the vulnerable library itself, we replace the missing version with the closest available version. 

\item \textbf{S2: Switching to Appropriate JDK/System.} 
We resolve most of the \textsc{JDK Mismatch} issues (117/123, $95.1\%$) and 8 \textsc{System Mismatch} issues ($100\%$) by aligning the environment. In most cases, the error logs explicitly indicate the required version or system (e.g., \textit{class file has wrong version 61.0, should be 55.0}). However, a subset of cases requires additional inference. For example, in the \textit{XStream} library, the error \textit{Class not found: LambdaMapper} does not directly mention JDK versions, but our analysis revealed it is caused by the removal of certain reflection capabilities in newer JDKs. Switching to an older JDK resolves the issue. 

\item \textbf{S3: Modifying Build Configuration Files.} 
We resolve all 22 \textsc{Other Library Incompatibility} cases ($100\%$) by ensuring the runtime classpath matches the environment expected by the exploit. These incompatibility issues are typically well-documented in the error logs (e.g., \textit{LogFactory class must be in class path}), providing clear guidance for resolution. 

\end{itemize}

However, the remaining 10 cases are marked as unresolvable due to the unavailability of essential artifacts. Specifically, in four versions, the vulnerable library is absent from the Maven Central Repository. In another six versions, the required JDK are no longer accessible (e.g.,\textit{Java 1.8.0-ea-b99+}).

\subsubsection{Library Incompatibility Issues}

Among 1,611 cases with incompatibility issues, such as \textsc{Symbol Resolution Failure}, we migrate 1,432 cases ($88.9\%$) and summarize four strategies:

\begin{itemize}[leftmargin=*]

\item \textbf{S4: Locating Equivalent Trace.} 
We migrate exploits for 1,067 versions where failures are caused by software refactoring, such as package renaming or method movement. Leveraging the diff files of the library, we identify the renamed classes or methods that preserve the original functionality in the target version. We then update the exploit to follow these equivalent traces. 

\item \textbf{S5: Adapting Argument Types.} 
We migrate exploits for 143 cases where the target API exists but requires different parameter types compared to the original exploit (e.g., from accepting a \textit{File} to an \textit{InputStream}). We resolve these by manually converting the data types before passing them to the API to comply with the target version constraints.  

\item \textbf{S6: Reconstructing Logic with Alternative Paths.} 
We migrate exploits for 132 cases where the API does not exist or the default trace is insufficient to trigger the vulnerability. In these scenarios, we reconstruct the exploit logic by combining other available APIs to achieve the same semantic goal.
For example, in \textit{Jinjava} (v2.0.2), the method \textit{newInterpreter()} is absent. We replace it by invoking the constructor and context retrieval functions to replicate the initialization logic.
Similarly, for certain \textit{XStream} versions where the default initialization fails to trigger the deserialization vulnerability, we initialize it with \textit{Sun14ReflectionProvider()} to bypass the restriction logic and reproduce the vulnerability. 

\item \textbf{S7: Adjusting Transitive Dependency Versions.} 
We resolve 90 cases where ``Not Found'' errors stem from missing transitive dependencies rather than the target library itself. For instance, \textit{`java.lang.NoClassDefFoundError: net/minidev/asm/FieldFilter'} indicates a mismatch in helper libraries. We resolve these by identifying and switching the transitive library to versions compatible with the target environment (\textit{accessors-smart:2.5.1}). 

\end{itemize}

However, for \textbf{179} cases (associated with 9 vulnerabilities), our migration attempts failed. The primary reason is the inability to extract clear replacement logic or equivalent behaviors from the diff files. 
In these scenarios, the target versions (typically older ones) fundamentally lacked the features or logical components required to trigger the vulnerability.
For instance, regarding \textit{CVE-2022-24897}, the reproduction of the vulnerability relies on a specific API to manipulate security configurations. However, in lower versions of the library, this API and the underlying security feature it controls are not yet implemented. Consequently, no valid migration path could be derived from the diffs, rendering the exploit not migratable.

\subsubsection{Exploit Failure}

Among the 672 cases where exploit reproduction failed due to semantic discrepancies or logic errors, we successfully migrate \textbf{299} cases (44.5\%). We summarize three strategies used to resolve these failures:

\begin{itemize}[leftmargin=*]

\item \textbf{S8: Adjusting Verification Assertions.} 
We migrate exploits for 69 versions where the vulnerability is triggered, but the reproduction behavior is not caught by the designed assertions. We resolved this by modifying assertions to align with the actual behaviors of the target version. 

\item \textbf{S9: Refining Payload Formats.} 
We migrate exploits for 139 versions where failures are caused by strict format validation in the target library. We resolved these by adding or removing specific fields in the payload to comply with the target version parsing requirements, ensuring the payload structure satisfies the internal checks while preserving the malicious logic. For instance, in \textit{json-java:20170516}, strict lexical analysis for the input JSON causes the exploit for \textit{CVE-2023-5072} to fail due to the presence of unsupported control characters. We migrate the exploit by removing the unrecognized sequence \textit{\textbackslash t\textbackslash 0} from the payload. 

\item \textbf{S10: Modifying Exploit Execution Traces.} 
We resolve 91 cases by altering the execution path of the exploit to bypass runtime constraints or errors. This involves pruning problematic branches or modifying the gadget chain.
For \textit{CVE-2019-16942} in \textit{jackson-databind:v2.2.4}, the exploit fails with \textit{`Conflicting getter definitions for property testOnBorrow'}. We resolve this by introducing a Jackson \textit{MixIn} with \textit{@JsonIgnore} to skip the problematic property validation, pruning the crashing branch, and making the exploit viable again.
For \textit{CVE-2022-25845} in \textit{fastjson:1.2.39}, the original exploit uses \textit{java.lang.Exception} as the entry point for the deserialization chain, which failed in this version. We successfully migrate it by switching the entry point to \textit{java.lang.Class}, which allows the subsequent gadget chain to execute correctly. 

\end{itemize}

We failed to migrate exploits in scenarios where the modification logic relied heavily on deep domain expertise rather than explicit code changes. 
For instance, payloads generated via fuzzing or gadgets involving complex deserialization mechanisms often lack clear correspondence in the source code. 
In these cases, relying solely on Diff analysis is insufficient and requires specialized domain knowledge or the assistance of existing exploit generation tools.

\subsubsection{Migration Cost Analysis}

Table~\ref{tab:strategies_horizontal} summarizes the average time cost and the number of modified lines of code (LoC) for each migration strategy. 
Strategies addressing environmental issues (S1--S3) incur the lowest overhead, typically requiring less than four minutes to resolve through simple version alignment or configuration adjustments. 
In contrast, handling library incompatibilities (S4--S7) involves more effort, with the required time largely correlating with the extent of code modifications needed for API adaptation. 
Resolving exploit failures (S8--S10) is the most time-consuming category, taking up to 55.1 minutes on average, which reflects the substantial effort required for comprehending the underlying logic and modifying the exploit accordingly.

\begin{table}[t] 
\centering
\caption{Migration Cost across All Strategies.}
\label{tab:strategies_horizontal}
\small
\setlength{\tabcolsep}{4.5pt} 
\resizebox{0.9\linewidth}{!}{%
\begin{tabular}{@{}l|ccc|cccc|ccc@{}}
\toprule
\textbf{Category} & \multicolumn{3}{c|}{\textbf{Environment}} & \multicolumn{4}{c|}{\textbf{Library Incompatibility}} & \multicolumn{3}{c}{\textbf{Exploit Failure}} \\ \midrule
\textbf{Strategy} & \textbf{S1} & \textbf{S2} & \textbf{S3} & \textbf{S4} & \textbf{S5} & \textbf{S6} & \textbf{S7} & \textbf{S8} & \textbf{S9} & \textbf{S10} \\ \midrule
\textbf{Time (min)} & 3.86 & 2.83 & 21.3 & 14.4 & 16.1 & 26.4 & 25.3 & 37.0 & 51.4 & 55.1 \\
\textbf{LoC (lines)} & 7.86 & 1.72 & 3.55 & 16.1 & 22.7 & 18.9 & 28.9 & 23.0 & 15.8 & 14.2 \\ \bottomrule
\end{tabular}
}

\end{table}

\mybox{
\textbf{Finding 4}: \rev{
    We identify \textbf{10 migration strategies} that facilitate exploit migration based on our manual migration, recovering \textbf{77.0\%} of failed cases and \textbf{raising the achievable recall to 96.1\%}.
    While strategies handling environment and compatibility issues are effective based on code changes, those addressing semantic logic divergences remain challenging.
    }
}

\section{Discussion}
\label{sec: discussion}

In this section, we present the implications of our findings, followed by an analysis of the threats to validity.

\subsection{Implications to Researchers and Developers}

We derive our implications from three perspectives: vulnerability databases, further researchers, and downstream developers.

\textbf{Vulnerability Databases should leverage exploits to complete affected version records for Java library vulnerabilities.} Current databases (e.g., NVD, GitHub Advisory) primarily rely on static metadata or manual reports to determine affected versions, which we found to be error-prone and incomplete. Our Findings 1 and 2 demonstrate that exploit-based assessment is feasible and effective. This implies that these databases can leverage community-contributed exploits to validate the full history of library versions. The practicality is evidenced by our cross-version execution, as we submit more than 1,400 missing affected versions to CPE.

\textbf{Researchers should advance automated exploit migration to handle complex scenarios.}
We provide the largest dataset for exploit migration. Based on our Findings 3 and 4, we identify the direction for future research on automated tools:

\begin{itemize}[leftmargin=*] \item \textbf{Validating Diff-based Strategies as a Strong Baseline:} Our study confirms that strategies guided by code changes (i.e., Diff-based) are highly effective for the majority of migration tasks, achieving a success rate of over 77\%. Future research should treat diff-based analysis as a baseline of their migration pipeline, as it efficiently handles straightforward API renames and signature updates without the need for expensive, heavyweight analysis. With exploit migration, the recall of exploit-based affected version identification reaches 96.1\% on our dataset, outperforming all evaluated vulnerability databases and tools.

\item \textbf{Implementing the Direction for Future Research:} 
Our proposed migration strategies demonstrate that exploit migration is feasible when structural or environment-related changes are explicitly captured. Nevertheless, our study also exposes the limitations of current strategies. In particular, cases involving semantic logic changes remain difficult to handle automatically, which opens an important research direction for intelligent and semantic-aware exploit migration.
\end{itemize}

\textbf{Downstream Developers should adopt a comprehensive and context-aware impact assessment strategy.} Our Finding 1 reveals that relying on a single vulnerability database is insufficient, as even the best-performing database exhibits inaccuracies. Developers should therefore cross-reference multiple data sources to minimize misjudgment. Furthermore, we find that exploitability is strictly tied to runtime contexts, such as specific JDK versions. Consequently, developers must incorporate environmental constraints into their verification process, moving beyond simple version matching to assess whether the vulnerability is actually exploitable in their specific deployment environment.



\subsection{Threats to Validity}

\textbf{External Threats.} The primary threat to the generalizability arises from the exploit collection process. We collect vulnerabilities from a widely used package management system, Maven, for evaluation. Vulnerability exploits in other ecosystems are not involved in our evaluation, which may limit the generalizability of the results across different ecosystems. In addition, while our dataset covers 76.33\% of the 5,889 vulnerabilities in terms of CWE coverage, the remaining 23.67\% are still not covered by our dataset. Using the 2025 CWE Top 25 list, we find that only three Top-25 CWEs are not covered. Specifically, CWE-352 and CWE-639 mainly belong to platforms outside of our scope, such as xxl-job, while CWE-122 often overlaps with the already covered CWE-787 in Java. Moreover, in-the-wild exploits are excluded. To address this, we note that in our researched exploits, Wu et al.~\cite{Wu2024Vision} have collected such examples.

An additional external threat to our evaluation arises from the difficulty of reliably reproducing exploits. This process is often challenged by issues such as incomplete or unreliable official exploit implementations. To mitigate this threat, we first filtered exploits based on established criteria and then conducted manual reproduction by experienced developers. The reproduction results were further cross-validated to ensure correctness and consistency.

\textbf{Internal Threats.} The collected exploits may not correspond directly to the vulnerabilities, and when constructing exploits based on vulnerability patches, irrelevant tests may interfere. To mitigate this, we assess whether the performance in the exploit reports aligns with the actual behavior to determine successful exploitation.

\textbf{Construct Threats.} 
Regarding the comparative assessment, the contents of vulnerability databases are dynamic and constantly updated, causing construct threats. To ensure a deterministic and fair comparison, we evaluate the capabilities of all data sources based on their snapshots as of June 2025. We provide the exact snapshot data in our replication package to support reproducibility.

\section{Related Work}
\label{sec: related}

\textbf{Vulnerable Version Assessment.} Assessing the affected versions of vulnerabilities is crucial for mitigating risks. Existing studies have proposed automated methods to identify affected versions~\cite{Dong2019Inconsistence, Bao2022Vszz, He2024Logs, Shi2023Web, Wu2024Vision, Dai2021Exploit, Jiang2023AEM, zou2024SyzBridge, Zhou2024Magneto}. Dong et al.~\cite{Dong2019Inconsistence} introduced VIEM to detect inconsistencies between the NVD database and CVE descriptions. Bao et al.~\cite{Bao2022Vszz} proposed V-SZZ to refine the set of versions impacted by vulnerabilities. Patches have also been utilized to detect affected versions~\cite{He2024Logs, Shi2023Web, Wu2024Vision}. He et al.~\cite{He2024Logs} leveraged developer logs and patches to identify affected versions. Wu et al.~\cite{Wu2024Vision} encode the criticality of vulnerable methods and statements into weighted IPDGs for accurate identification. Shi et al.~\cite{Shi2023Web} introduced a patch-based method to determine affected versions of web vulnerabilities. \textbf{Vulnerability exploits} play a critical role in assessing affected versions. Dai et al.~\cite{Dai2021Exploit} proposed a fuzzing-based approach to guide the mutation of exploit inputs outside the reference version. Jiang et al.~\cite{Jiang2023AEM} introduced AEM to align exploits between reference versions and other kernel versions. Although these studies leverage exploits to confirm affected versions, the applicability of exploit-based assessment remains unexplored.

\textbf{Library Vulnerability Exploits.} 
A primary objective of our study is to construct a vulnerability exploit dataset. To this end, we first examine existing datasets in prior research~\cite{Dai2021Exploit,Yang2023oneday,zirui2024exploiting,gao2025exploit,chen2025mitigation,Wu2024Vision}. Dai et al.~\cite{Dai2021Exploit} collected 30 real-world CVEs spanning 470 software versions and evaluated the reliability of each exploit input. However, the CVEs were randomly selected, which reduced the transparency. A similar limitation is observed in the dataset constructed by Yang et al. ~\cite{Yang2023oneday}. Although Chen et al.~\cite{zirui2024exploiting} constructed datasets targeting Java library vulnerabilities, they did not disclose the details of the data collection processes, thus compromising the transparency of the coverage.  The dataset of Wu et al.~\cite{Wu2024Vision} contains fixed verification tests rather than exploits in their dataset.

\textbf{Vulnerability Knowledge Enhancement.}Researchers have extensively analyzed the inconsistency of CVSS scores~\cite{Anwar2022CVSS, Croft2022CVSS, wunder2024shedding} and developed predictive models for exploitability ~\cite{Chen2019exploit, miranda2023patch}. Beyond assessment, accurate localization of vulnerabilities is critical. Significant efforts have been dedicated to automated patch identification ~\cite{Nguyen2022Commit,Sabetta2024fix, Tan2021fix, Xu2022Patch}, affected library identification~\cite{Wu2024SCA, chen2020automated, chen2025identifying}, and vulnerable version assessment~\cite{Wu2024Vision, Shi2023Web, He2024Logs, Bao2022Vszz}.

\section{Conclusion}
\label{sec: conclusion}

In this paper, we conduct the first large-scale empirical study to investigate the cross-version applicability of exploits, challenging the prevailing limitation that exploits are version-specific. By constructing a comprehensive dataset, we demonstrate that disclosed exploits exhibit high performance, achieving an 83.0\% recall and 99.3\% precision. These results highlight the role of exploits in verifying affected versions. Furthermore, our root cause analysis reveals that the primary barriers to exploit migration are breaking changes during library evolution. Through our derived migration strategies, 77.1\% of failed cases on vulnerable versions can be successfully recovered, increasing the recall of affected version identification to 96.1\%. We hope that our findings and the released dataset serve as a foundational benchmark for future research, facilitating the broader adoption of exploit-driven vulnerable version assessment.

\section{Data Availability}
The original exploits, migrated exploits, and source code are publicly available in our replication package~\cite{Package}.

\section*{Acknowledgement}
This research is supported by National Key R\&D Program of China (No. 2024YFB4506400) and the CCF-Huawei Populus Grove Fund. We also thank the anonymous reviewers for their insightful comments and suggestions.

\balance
\bibliographystyle{ACM-Reference-Format}
\bibliography{main}

\end{document}